\newif\ifpreprint
	\preprintfalse
 	\preprinttrue

\ifpreprint
	\documentstyle[aas2pp4,epsf,twoside,fleqn]{article}
	\newcommand{\noi}	{\noindent}
\else
	\documentstyle[12pt,epsf,aasms4]{article}
	\newcommand{\noi}	{}
\fi

\newcommand{\bea}	{\begin{array}}
\newcommand{\eea}	{\end{array}}
\newcommand{\beq}	{\begin{equation}}
\newcommand{\eeq}	{\end{equation}}
\newcommand{\ben}	{\begin{eqnarray}}
\newcommand{\een}	{\end{eqnarray}}
\newcommand{\bsq}	{\begin{mathletters}}
\newcommand{\esq}	{\end{mathletters}}

\newcommand{\B}[1]	{\mbox{\boldmath$#1$}}

\newcommand{\bvel}	{\B{v}}

\newcommand{\D}		{{\rm d}}

\newcommand{\Rcr}	{R_{\rm\scriptscriptstyle CR}}
\newcommand{\Rolr}	{R_{\rm\scriptscriptstyle OLR}}
\newcommand{\Ro}	{R_0}
\newcommand{\Rb}	{R_{\rm b}}
\newcommand{\vc}	{v_{\rm c}}
\newcommand{\vo}	{v_0}
\newcommand{\vterm}	{v_{\rm term}}
\newcommand{\volr}	{v_{\rm\scriptscriptstyle OLR}}
\newcommand{\half}	{\case{1}{2}}

\newcommand{\fuv}	{\mbox{$f(u,v)$}}
\newcommand{\Ob}	{\Omega_{\rm b}}
\newcommand{\op}	{\omega_\phi}
\newcommand{\oR}	{\omega_R}
\newcommand{\Oo}	{\Omega_0}

\newcommand{\SgrA}	{Sgr\,A$^{\!\star}$}

\newcommand{\msun}	{\mbox{${\rm M}_\odot$}}
\newcommand{\kpc}	{\mbox{\,kpc}}

\newcommand{\kms}	{\mbox{\,km\,s$^{-1}$}}
\newcommand{\kmskpc}	{\mbox{\,km\,s$^{-1}$\,kpc$^{-1}$}}
\newcommand{\mkmskpc}	{\mbox{km\,s$^{-1}$\,kpc$^{-1}$}}

\newcommand{\etal}	{\mbox{et~al.}}
\newcommand{\eqn}[1]	{equation\ (\ref{#1})}

\newcommand{\eqb}[1]	{(\ref{#1})}
\newcommand{\Sec}[1]	{\S\ref{sec:#1}}
\newcommand{\Fig}[1]	{Figure\ \ref{fig:#1}}
\newcommand{\fig}[1]	{Fig.\ \ref{fig:#1}}
\newcommand{\Tab}[1]	{Table\ \ref{tab:#1}}

\ifpreprint
\else
	\lefthead{Walter Dehnen}
	\righthead{The Pattern Speed of the Galactic Bar}
\fi

\begin{document}

\ifpreprint \thispagestyle{empty} \fi

\title{The Pattern Speed of the Galactic Bar}
\author{Walter Dehnen}
\affil{	Theoretical Physics, 1 Keble Road, Oxford OX1 3NP, United Kingdom; and\\
	Max-Planck Institut f.\ Astronomie, K\"onigstuhl, D-69117 Heidelberg,
	Germany; dehnen@mpia-hd.mpg.de
}

\begin{abstract} \noi
Most late-type stars in the solar neighborhood have velocities similar to the
local standard of rest (LSR), but there is a clearly separated secondary
component corresponding to a slower rotation and a mean outward motion. Detailed
simulations of the response of a stellar disk to a central bar show that such a
bi-modality is expected from outer-Lindblad resonant scattering. When
constraining the run of the rotation curve by the proper motion of \SgrA\ and
the terminal gas velocities, the value observed for the rotation velocity
separating the two components results in a value of $(53\pm3)\kmskpc$ for the
pattern speed of the bar, only weakly dependent on the precise values for $\Ro$
and bar angle $\phi$.
\end{abstract}

%%%%%%%%%%%%%%%%%%%%%%%%%%%%%%%%%%%%%%%%%%%%%%%%%%%%%%%%%%%%%%%%%%%%%%%%%%%%%%%%
\section{Introduction} \label{sec:intro} \noi
%%%%%%%%%%%%%%%%%%%%%%%%%%%%%%%%%%%%%%%%%%%%%%%%%%%%%%%%%%%%%%%%%%%%%%%%%%%%%%%%
Even though we now know beyond reasonable doubt that the Milky Way is barred,
the structure of the bar, its orientation and pattern speed are still subject to
substantial debate, mainly because of the edge-on view and dust obscuration.
Consequently, the properties of the bar have been inferred rather indirectly
from IR photometry (\cite{bs91}; \cite{dw95}; \cite{bgs97}), asymmetries in the
distribution or magnitude of stars (\cite{wc92}; \cite{wb92}; \cite{nw97};
\cite{st95}; \cite{sev96}; \cite{st97}), and gas velocities (\cite{lb80}; 
\cite{b91}; \cite{eg99}; \cite{ws99}). While all these studies are based on
observations of the inner Galaxy itself, it is the goal of this letter to
constrain the properties of the bar by its effect on the stellar velocity
distribution observed in the solar neighborhood.

% \placefigure{fig:fuvobs}

%%%%%%%%%%%%%%%%%%%%%%%%%%%%%%%%%%%%%%%%%%%%%%%%%%%%%%%%%%%%%%%%%%%%%%%%%%%%%%%%
\section{The Local Stellar Velocity Distribution} \label{sec:dist} \noi
%%%%%%%%%%%%%%%%%%%%%%%%%%%%%%%%%%%%%%%%%%%%%%%%%%%%%%%%%%%%%%%%%%%%%%%%%%%%%%%%
In contrast to the structure of the inner Galaxy, the local stellar velocity
distribution can be observed to much greater detail than for any external
galaxy. \Fig{fuvobs} shows the distribution \fuv\ in radial ($u$) and rotational
($v$) velocities (relative to the LSR) that has been inferred from the Hipparcos
data (\cite{hipcat}) of 6\,018 late-type stars (\cite{deh98} \& 1999a).

This distribution shows a lot of structure, most of which is real (noise has
been suppressed). Apart from an overall smooth background and a {\em
low-velocity\/} region (solid ellipse in \fig{fuvobs}), within which $f$ has
several peaks associated with moving groups, there is also an {\em
intermediate-velocity\/} structure (broken ellipse) containing $\sim15$\% of the
late-type stars but hardly any early-type stars, which mainly populate the
moving groups (\cite{deh98}; \cite{ch98}; \cite{asi99}; \cite{shc99}). This
secondary component, which consists predominantly of stars moving outwards
($u<0$) and is clearly separated from the low-velocity peak, was only vaguely
recognizable in pre-Hipparcos data sets; the resulting mean outward motion of
stars with $v\la-30\kms$ is also known as $u$-anomaly.

\ifpreprint
    \begin{figure}[t]
	\centerline{\epsfxsize=85mm\epsfbox[20 17 424 308]{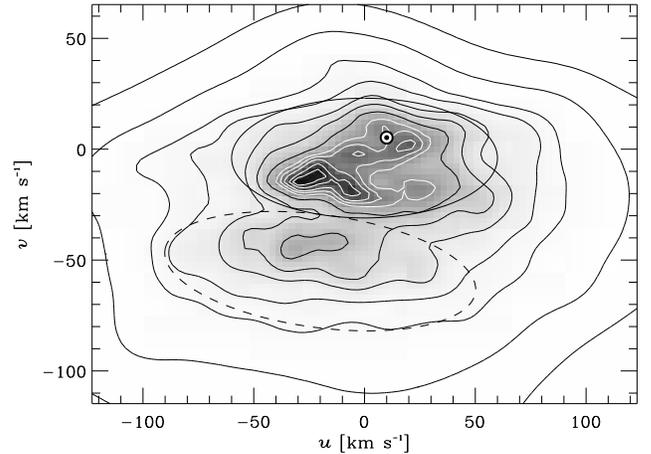}}
	\caption[]{\footnotesize
	The distribution \fuv\ inferred from Hipparcos data for late-type stars
	(3\,527 main-sequence stars with $\bv\ge0.6$ and 2\,491 mainly late-type
	non-main-sequence stars, high-velocity stars excluded, see
	\cite{deh98}). $u$ and $v$ denote the velocities towards $\ell=0^\circ$
	and $\ell=90^\circ$ w.r.t.\ the LSR measured by Dehnen \& Binney (1998);
	$\odot$ indicates the solar velocity. Samples of early-type stars
	contribute almost exclusively to the low-velocity region (solid
	ellipse). The region of intermediate velocities (broken ellipse) is
	mainly represented by late-type stars, $\sim15\%$ of which fall into
	this region. \label{fig:fuvobs} }
    \end{figure}
\fi     %preprint

%%%%%%%%%%%%%%%%%%%%%%%%%%%%%%%%%%%%%%%%%%%%%%%%%%%%%%%%%%%%%%%%%%%%%%%%%%%%%%%%
\section{What Causes this Bi-Modality?} \label{sec:cause} \noi
%%%%%%%%%%%%%%%%%%%%%%%%%%%%%%%%%%%%%%%%%%%%%%%%%%%%%%%%%%%%%%%%%%%%%%%%%%%%%%%%
\ifpreprint
Attempts to explain this seeming anomaly by non-equilibrium dynamics, like the
\else
Attempts to explain this seeming anomaly by non-equi\-librium dynamics, like the
\fi
dispersion of a stellar cluster or the merger of a satellite, fail because of
the exclusively late stellar type, disk-like kinematics ($\bvel$ nearly as LSR),
and rather high metallicity (\cite{rab98}) of the intermediate-velocity stars.
Thus, dynamical equilibrium must account both for the mean outward motion of the
intermediate-velocity stars and for their clear separation from the low-velocity
stars. The first fact rules out any axisymmetric equilibrium, while the second
strongly hints towards an orbital resonance.

%%%%%%%%%%%%%%%%%%%%%%%%%%%%%%%%%%%%%%%%%%%%%%%%%%%%%%%%%%%%%%%%%%%%%%%%%%%%%%%%
\subsection{The Bar's Outer Lindblad Resonance} \noi
%%%%%%%%%%%%%%%%%%%%%%%%%%%%%%%%%%%%%%%%%%%%%%%%%%%%%%%%%%%%%%%%%%%%%%%%%%%%%%%%
Such a resonance is provided by the Galactic bar: the outer Lindblad resonance
(OLR), which occurs when
\beq \label{reso}
	\Ob = \op + \half \oR.
\eeq
Here, $\op$ and $\oR$ are the azimuthal and radial orbital frequencies, while
$\Ob$ is the pattern speed of the bar. After each radial lobe, a star on a
resonant orbit is at the same position w.r.t.\ bar, and hence, is always pushed
in the same direction forcing it onto a different orbit. According to the
modeling of the gas motions in the inner Galaxy (\cite{eg99}; \cite{ws99}), the
radius $\Rolr$ corresponding to the OLR of circular orbits is not far from
$\Ro$.

Orbits inside (outside) the OLR, i.e.\ with $\op+\half\oR>\Ob$ ($<\Ob$), are
elongated perpendicular (parallel) to the bar's major axis and are moving on
average outwards (inwards) for bar angles $\phi\in[0^\circ,90^\circ]$. The
bar angle is defined as the azimuth of the Sun w.r.t.\ the bar's major axis and,
according to IR photometry, lies in the range $15^\circ$ to $45^\circ$.

Thus, the OLR of the Galactic bar can naturally explain the low- versus
intermediate-velocity bi-modality of \fuv, provided the Sun is {\em outside} the
OLR (\cite{deh99a}). In this picture, the depression between the modes of \fuv\
is due to exactly resonant orbits, and from the observed $v$-velocity of
\beq \label{new}
	\volr = (-31 \pm 3) \kms
\eeq
for the saddle point between the two modes, we can estimate that $\Ro-\Rolr
\simeq|\volr/\vo|\Ro\simeq1\kpc$ (for a flat rotation curve). Thus, only stars
with epicycle amplitudes $\ga1\kpc$, i.e.\ predominantly late-type stars, can
visit us from inside the OLR, in agreement with the absence of early-type stars
in the secondary component.

This proposal is in line with the ideas of Raboud \etal\ (1998), who observed a
high metallicity and outward mean motion for disk stars with $v<{-}30\kms$ and
interpreted this as evidence for the Galactic bar, since in Fux's (1997) 
$N$-body simulations of the Milky Way $\bar{u}<0$ at the solar position. 
However, these authors did not relate the effect to the OLR nor were they able, 
for lack of resolution both in the data and the model, to make detailed 
quantitative investigations.

%%%%%%%%%%%%%%%%%%%%%%%%%%%%%%%%%%%%%%%%%%%%%%%%%%%%%%%%%%%%%%%%%%%%%%%%%%%%%%%%
\subsection{The Response of the Disk to a Stirring Bar} \label{sec:simul} \noi
%%%%%%%%%%%%%%%%%%%%%%%%%%%%%%%%%%%%%%%%%%%%%%%%%%%%%%%%%%%%%%%%%%%%%%%%%%%%%%%%
In order to quantify the behavior of a warm stellar disk in presence of a
central bar, I performed numerical simulations in which a bar is grown in an
initially axisymmetric exponential disk. These simulations, presented in
detail in Dehnen (1999b), use a technique, which is similar to that of
Vauterin \& Dejonghe (1997) and allows for very high resolution.

The simulations show that a second mode appears in \fuv\ at many positions in
the Galactic disk, provided the velocity dispersion is high enough. The velocity
$\volr$ of the division line between the modes depends essentially on four 
parameters: (1) our distance to the OLR, quantified by the ratio $\Rolr/\Ro$;
(2) the bar angle $\phi$; the (3) shape and (4) normalization of the rotation
curve, quantified by $\beta\equiv\D\ln\vc/\D\ln R$ and $\vo\equiv\vc(\Ro)$. The
strength of the bar, its morphology, and the details of the stellar DF hardly
affect the velocity $\volr$, but may change the strength of the secondary
component.

% \placefigure{fig:simul}

The simulations used rotation curves $\vc\propto R^\beta$ and a simple
quadrupole to model the non-axisymmetric contribution of the bar potential.
\Fig{simul} shows the resulting \fuv\ for $\beta=0$, $\phi=25^\circ$, and 
three different values for $\Ob$, which is related to $\Rolr$ by
\beq \label{RoR}
	{\Rolr\over\Ro} = \left({\Oo\over\Ob}
		\left[1+\sqrt{1+\beta\over2}\right]\right)^{1/(1-\beta)}
\eeq
with $\Oo\equiv\vo/\Ro$. The secondary component is weaker and occurs at larger
$|v|$ the farther the OLR is from the Sun. For $15^\circ\le\phi\le45^\circ$,
$-0.2\le\beta\le0.2$ and $0.8\le\Rolr/\Ro\le0.95$, $\volr$ is well approximated
by
\beq \label{fit}
\volr \approx a {1+\beta\over1-\beta} \left[\vo - 
			{\Ro\,\Ob\over1+\sqrt{(1+\beta)/2}} \right]
		 - (b + c\,\beta)\,\vo,
\eeq
where the coefficients $a$, $b$, and $c$ are given in \Tab{fit}.

% \placetable{tab:fit}

\ifpreprint
   \begin{figure}[t]
        \centerline{\epsfxsize=60mm \epsfbox[5 11 166 268]{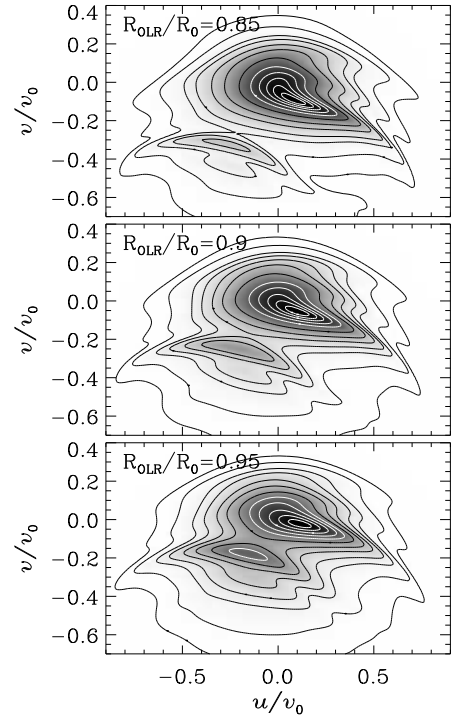}}
        \caption[]{\footnotesize
	Simulated \fuv\ after the growth of a central bar. The initial
	axisymmetric disk has scale length $0.33\Ro$ and velocity dispersion
	reminiscent of the old stellar disk. The rotation curve is assumed flat
	and the bar angle is fixed at $\phi=25^\circ$. The pattern speed
	decreases from top to bottom panel. \label{fig:simul} }
   \end{figure}

   \begin{table}[b]
        \caption[]{Best-fit values for $(a,b,c)$ in \eqn{fit}\label{tab:fit}}
        \begin{tabular}{l@{$\qquad$}r@{.}l@{$\qquad$}r@{.}l@{$\qquad$}r@{.}l}
                                        \\[-2ex] \hline\hline \\[-2.5ex]
        $\phi$ & \multicolumn{2}{c}{$a$} &
        \multicolumn{2}{c}{$b$} & \multicolumn{2}{c}{$c$}       \\ \hline
                                                                \\[-2.3ex]
        15$^\circ$      &1&3549 &0&0761 &0&1362 \\
        20$^\circ$      &1&2686 &0&0642 &0&1120 \\
        25$^\circ$      &1&2003 &0&0526 &0&0892 \\
        30$^\circ$      &1&1424 &0&0406 &0&0711 \\
        35$^\circ$      &1&0895 &0&0298 &0&0538 \\
        40$^\circ$      &1&0420 &0&0200 &0&0423 \\
        45$^\circ$      &1&0012 &0&0103 &0&0316 \\              \hline\hline
        \end{tabular}
    \end{table}

    \begin{table*}[t]
	\centering
        \caption[]{\centering Adding the new constraint: fitting $\vterm$
                for $|\ell|>30^\circ$, $\mu_\ell$ of \SgrA, and $\volr$.
                \label{tab:new}}
        \begin{tabular}{l|r@{$\,\pm\,$}lr@{$\,\pm\,$}lc} 
        \multicolumn{6}{c}{} \\[-2ex] \hline\hline \multicolumn{6}{c}{} \\[-2ex]
        \multicolumn{6}{c}{$\Ro=7$, $\vo=198\pm5.6$, 
                        $\beta={-}0.147\pm0.025$}		\\ \hline
        $\phi$ &\multicolumn{2}{c}{$\Ob$}       & 
                \multicolumn{2}{c}{$\Rolr/\Ro$} &  
                \multicolumn{1}{c}{$\Rcr/\Ro$}                  \\ \hline
        15$^\circ$ & 51.4&1.4 & 0.920&0.011 & 0.594 \\
        20$^\circ$ & 52.1&1.5 & 0.909&0.012 & 0.586 \\
        25$^\circ$ & 52.8&1.5 & 0.898&0.012 & 0.579 \\
        30$^\circ$ & 53.7&1.5 & 0.886&0.013 & 0.572 \\
        35$^\circ$ & 54.5&1.6 & 0.874&0.013 & 0.564 \\
        40$^\circ$ & 55.3&1.6 & 0.863&0.013 & 0.557 \\
        45$^\circ$ & 56.2&1.6 & 0.851&0.013 & 0.549 \\
        \hline\hline \multicolumn{6}{c}{} \\[-2ex]
        \multicolumn{6}{c}{$\Ro=7.5$, $\vo=212\pm6$,
                        $\beta={-}0.080\pm0.024$}		\\ \hline
        $\phi$ &\multicolumn{2}{c}{$\Ob$}       & 
                \multicolumn{2}{c}{$\Rolr/\Ro$} &  
                \multicolumn{1}{c}{$\Rcr/\Ro$}                  \\ \hline
        15$^\circ$ & 50.8&1.4 & 0.939&0.010 & 0.581 \\
        20$^\circ$ & 51.5&1.4 & 0.928&0.011 & 0.575 \\
        25$^\circ$ & 52.2&1.4 & 0.917&0.011 & 0.568 \\
        30$^\circ$ & 52.9&1.5 & 0.905&0.011 & 0.560 \\
        35$^\circ$ & 53.7&1.5 & 0.893&0.012 & 0.553 \\
        40$^\circ$ & 54.4&1.5 & 0.882&0.012 & 0.546 \\
        45$^\circ$ & 55.2&1.6 & 0.870&0.012 & 0.539 \\
        \hline\hline
        \end{tabular}
                \hspace{1.5cm}
        \begin{tabular}{l|r@{$\,\pm\,$}lr@{$\,\pm\,$}lc} 
        \multicolumn{6}{c}{} \\[-2ex] \hline\hline \multicolumn{6}{c}{} \\[-2ex]
        \multicolumn{6}{c}{$\Ro=8$, $\vo=227\pm6.4$,
                        $\beta={-}0.020\pm0.022$}		\\ \hline
        $\phi$ &\multicolumn{2}{c}{$\Ob$}       & 
                \multicolumn{2}{c}{$\Rolr/\Ro$} &  
                \multicolumn{1}{c}{$\Rcr/\Ro$}                  \\ \hline
        15$^\circ$ & 50.5&1.4 & 0.954&0.010 & 0.567 \\
        20$^\circ$ & 51.1&1.4 & 0.943&0.010 & 0.561 \\
        25$^\circ$ & 51.8&1.4 & 0.932&0.010 & 0.554 \\
        30$^\circ$ & 52.5&1.4 & 0.920&0.010 & 0.547 \\
        35$^\circ$ & 53.2&1.5 & 0.908&0.011 & 0.540 \\
        40$^\circ$ & 53.8&1.5 & 0.897&0.011 & 0.533 \\
        45$^\circ$ & 54.5&1.5 & 0.885&0.011 & 0.526 \\
        \hline\hline \multicolumn{6}{c}{} \\[-2ex]
        \multicolumn{6}{c}{$\Ro=8.5$, $\vo=241\pm6.8$,
                        $\beta=0.033\pm0.021$}             \\ \hline
        $\phi$ &\multicolumn{2}{c}{$\Ob$}       & 
                \multicolumn{2}{c}{$\Rolr/\Ro$} &  
                \multicolumn{1}{c}{$\Rcr/\Ro$}                  \\ \hline
        15$^\circ$ & 50.4&1.4 & 0.967&0.009 & 0.552 \\
        20$^\circ$ & 51.0&1.4 & 0.956&0.009 & 0.546 \\
        25$^\circ$ & 51.6&1.4 & 0.944&0.009 & 0.539 \\
        30$^\circ$ & 52.2&1.4 & 0.932&0.010 & 0.532 \\
        35$^\circ$ & 52.8&1.4 & 0.921&0.010 & 0.526 \\
        40$^\circ$ & 53.5&1.5 & 0.909&0.010 & 0.519 \\
        45$^\circ$ & 54.1&1.5 & 0.898&0.010 & 0.513 \\
        \hline\hline 
        \end{tabular}
    \end{table*}
\fi	%preprint

%%%%%%%%%%%%%%%%%%%%%%%%%%%%%%%%%%%%%%%%%%%%%%%%%%%%%%%%%%%%%%%%%%%%%%%%%%%%%%%%
\section{Implications for the Inner Galaxy} \label{sec:impl} \noi
%%%%%%%%%%%%%%%%%%%%%%%%%%%%%%%%%%%%%%%%%%%%%%%%%%%%%%%%%%%%%%%%%%%%%%%%%%%%%%%%
We now {\em assume\/} that the observed low- versus intermediate-velocity
bi-modality of $f(u,v)$ is caused by scattering off the OLR of the Galactic
bar. Then the observed velocity $\volr$ \eqb{new} constitutes a new constraint
on the structure of the inner Galaxy, the implications of which we will
now investigate.

%%%%%%%%%%%%%%%%%%%%%%%%%%%%%%%%%%%%%%%%%%%%%%%%%%%%%%%%%%%%%%%%%%%%%%%%%%%%%%%%
\subsection{Other Constraints} \noi
%%%%%%%%%%%%%%%%%%%%%%%%%%%%%%%%%%%%%%%%%%%%%%%%%%%%%%%%%%%%%%%%%%%%%%%%%%%%%%%%
Unfortunately, the value of $\volr$ does not depend on just one, but several 
parameters. Thus, in order to understand the implications of the observed value,
we must combine the new constraint with other, independent constraints. 

Determinations of the proper motion of \SgrA, the radio source associated with
the Galactic center, yield (in \mkmskpc) $\mu_\ell=-28.0\pm1.7$
(\cite{reid99}) and $\mu_\ell=-29.3\pm0.9$ (\cite{bs99}), while the
latitudinal proper motion $\mu_b$ is negligible. \SgrA\ is generally thought to
be associated with a black hole of mass $\sim 2.6\times10^6\msun$
(\cite{eg97}, \cite{ghez98}), which because of energy equipartition must be
(nearly) at rest w.r.t.\ its surroundings. Thus, the observed proper motion is
entirely due to the reflex of our own rotation around the Galaxy, and we 
obtain
\beq \label{Om}
	\Oo = (29.0 \pm 0.8) \kmskpc - v_\odot / \Ro,
\eeq
where $v_\odot=(5.25\pm0.62)\kms$ is the Sun's $v$-motion w.r.t.\ the LSR
(\cite{db98}).

In order to constrain the shape of the rotation curve, i.e.\ the parameter
$\beta$, we use the terminal gas velocities inferred by Malhotra (1994, 1995)
from various observations in H\,{\sc i} and CO. For axisymmetric gas motions,
the terminal velocity is related to the rotation curve by
\beq \label{vterm}
	|\vterm| = \vc(R_0|\sin\ell|) - \vc(R_0)|\sin\ell|,
\eeq
while non-axisymmetric effects are likely to add small deviations to this
relation.

% \placetable{tab:new}

%%%%%%%%%%%%%%%%%%%%%%%%%%%%%%%%%%%%%%%%%%%%%%%%%%%%%%%%%%%%%%%%%%%%%%%%%%%%%%%%
\subsection{Adding the New Constraint} \noi
%%%%%%%%%%%%%%%%%%%%%%%%%%%%%%%%%%%%%%%%%%%%%%%%%%%%%%%%%%%%%%%%%%%%%%%%%%%%%%%%
We now put the various constraints together by (i) assuming (reasonable) values 
for $\Ro$ and $\phi$, (ii) computing $\vo=\Oo\Ro$ using \eqb{Om}, (iii) fitting 
$\beta$ to the observed terminal velocity curve, and (iv) deriving the bar's 
pattern speed $\Ob$ from the constraint on $\volr$ whereby using the 
approximation \eqb{fit}. Note that the bar angle becomes important only at 
stage (iv), i.e.\ $\vo$ and $\beta$ are already fixed once $\Ro$ is chosen.

For $\Ro=7$, 7.5, 8, and 8.5\kpc\ and values of $\phi$ in the range 15$^\circ$
to 45$^\circ$, which is required for the bar's morphology to be both similar to
that of external galaxies and consistent with IR photometry, \Tab{new} lists the
results of this procedure when fitting $\vterm$ at $|\ell|>30^\circ$ (the
results hardly depend to this limit). 

An inspection of the table reveals that the derived value for $\Ob$ is only
weakly dependent on the assumed values for $\Ro$ and $\phi$: $\Ob$ varies by
about 10\% with larger $\phi$ corresponding to larger $\Ob$. Owing to this
weak dependence, the pattern speed of the bar is rather tightly constrained even
without accurate knowledge of $(\Ro,\phi)$:
\beq \label{Ob}
	\Ob = (53\pm3)\kmskpc.
\eeq
\Tab{new} also lists estimates for $\Rolr/\Ro$ obtained from \eqn{RoR} and for
$\Rcr/\Ro$ obtained from the terminal velocities by the relation
\beq
	(\Ob-\Oo)\Ro = {\vterm(\ell)\over\sin\ell}\bigg|_{|\sin\ell|=\Rcr/\Ro}.
\eeq
An error is not given, but the uncertainty is about 10\%.

Amongst the three constraints used to derive $\Ob$ the biggest uncertainty comes
from the proper motion of \SgrA\ (and its interpretation as reflex of the solar
motion). If in \eqn{Om} one uses the value reported by Reid \etal, instead of a
weighted mean with Backer \& Sramek's measurement, the resulting $\Ob$ is about
2\% smaller, while $\Rcr/\Ro$ hardly changes.

The estimate \eqb{Ob} is also subject to systematic uncertainties, due mainly to
the idealized model. Future simulations which allow for a more general Galactic
rotation curve and bar potential and are fitted simultaneously to the gas motion
of the entire inner Galaxy and the local stellar velocity distribution are
likely to reduce these uncertainties considerably.

%%%%%%%%%%%%%%%%%%%%%%%%%%%%%%%%%%%%%%%%%%%%%%%%%%%%%%%%%%%%%%%%%%%%%%%%%%%%%%%%
\section{Conclusion} \label{sec:conc} \noi
%%%%%%%%%%%%%%%%%%%%%%%%%%%%%%%%%%%%%%%%%%%%%%%%%%%%%%%%%%%%%%%%%%%%%%%%%%%%%%%%
The simulations presented in Dehnen (1999b) and briefly reported in \Sec{simul}
show that the bar's OLR {\em inevitably\/} creates a bi-modality 
in \fuv\ over a wide range of positions in the Galactic disc. The low- versus
intermediate-velocity bi-modality in \fuv\ observed for late-type stars
(\fig{fuvobs}), has the same characteristics as these OLR features and is thus
best identified as such.

This in turn allows the observed $v$-velocity seperating the two modes to be
used as a new constraint on the parameters of the inner Galaxy, notably the
rotation curve, the bar's pattern speed and projection angle. In conjunction
with constraints for the rotation curve from the terminal gas velocities and the
proper motion of \SgrA, this results in a value for the bar's pattern speed of
$(53\pm3)\kmskpc$. The systematic uncertainty is likely similar in size to the
internal error; both shall be reduced in more elaborate future modeling.

The values for the corotation radius implied by this pattern speed and the
terminal gas velocities are between 0.5 and 0.6 times $\Ro$, i.e.\ larger than
the upper limit of 0.5 given by Englmaier \& Gerhard (1999), but consistent with
Weiner \& Sellwood (1999). Given that IR photometry suggests a bar length $\Rb$
of about $0.45\Ro$, this implies a ratio $\Rcr/\Rb$ of about $1.25\pm0.2$, in
agreement with the few estimates for external galaxies (\cite{mk95}; \cite{li96};
\cite{ger99}) and simulations (\cite{at92}).

%%%%%%%%%%%%%%%%%%%%%%%%%%%%%%%%%%%%%%%%%%%%%%%%%%%%%%%%%%%%%%%%%%%%%%%%%%%%%%%%
%
% TABLES
%
%%%%%%%%%%%%%%%%%%%%%%%%%%%%%%%%%%%%%%%%%%%%%%%%%%%%%%%%%%%%%%%%%%%%%%%%%%%%%%%%
\ifpreprint \relax \fi
%%%%%%%%%%%%%%%%%%%%%%%%%%%%%%%%%%%%%%%%%%%%%%%%%%%%%%%%%%%%%%%%%%%%%%%%%%%%%%%%
%
% REFERENCES using LaTeX's thebibliography environment
%
%%%%%%%%%%%%%%%%%%%%%%%%%%%%%%%%%%%%%%%%%%%%%%%%%%%%%%%%%%%%%%%%%%%%%%%%%%%%%%%%
\ifpreprint
  \def\thebibliography#1{\subsection*{References}
    \list{\null}{\leftmargin 1.2em\labelwidth0pt\labelsep0pt\itemindent -1.2em
    \itemsep0pt plus 0.1pt
    \parsep0pt plus 0.1pt
    \parskip0pt plus 0.1pt
    \usecounter{enumi}}
    \def\refpar{\relax}
    \def\newblock{\hskip .11em plus .33em minus .07em}
    \sloppy\clubpenalty4000\widowpenalty4000
    \sfcode`\.=1000\relax
    \footnotesize}
  \def\endthebibliography{\endlist}
\fi

%%%%%%%%%%%%%%%%%%%%%%%%%%%%%%%%%%%%%%%%%%%%%%%%%%%%%%%%%%%%%%%%%%%%%%%%%%%%%%%%
%
% FIGURE CAPTIONS
%
%%%%%%%%%%%%%%%%%%%%%%%%%%%%%%%%%%%%%%%%%%%%%%%%%%%%%%%%%%%%%%%%%%%%%%%%%%%%%%%%
\ifpreprint \relax \else
\clearpage \onecolumn

\begin{figure}[h]\caption[]{
	The distribution \fuv\ inferred from Hipparcos data for late-type stars
	(3\,527 main-sequence stars with $\bv\ge0.6$ and 2\,491 mainly late-type
	non-main-sequence stars, high-velocity stars excluded, see
	\cite{deh98}). $u$ and $v$ denote the velocities towards $\ell=0^\circ$
	and $\ell=90^\circ$ w.r.t.\ the LSR measured by Dehnen \& Binney (1998);
	$\odot$ indicates the solar velocity. Samples of early-type stars
	contribute almost exclusively to the low-velocity region (solid
	ellipse). The region of intermediate velocities (broken ellipse) is
	mainly represented by late-type stars, $\sim15\%$ of which fall into
	this region.  \label{fig:fuvobs} }
\end{figure}

\begin{figure}[h]\caption[]{
	Simulated \fuv\ after the growth of a central bar. The initial
	axisymmetric disk has scale length $0.33\Ro$ and velocity dispersion
	reminiscent of the old stellar disk. The rotation curve is assumed flat
	and the bar angle is fixed at $\phi=25^\circ$. The pattern speed
	decreases from top to bottom panel.
	\label{fig:simul} }
\end{figure}

\vfill

%%%%%%%%%%%%%%%%%%%%%%%%%%%%%%%%%%%%%%%%%%%%%%%%%%%%%%%%%%%%%%%%%%%%%%%%%%%%%%%%
%
% FIGURES 
%
%%%%%%%%%%%%%%%%%%%%%%%%%%%%%%%%%%%%%%%%%%%%%%%%%%%%%%%%%%%%%%%%%%%%%%%%%%%%%%%%

% FIGURE 1
	\clearpage
	\centerline{\epsfxsize=100mm\epsfbox[20 17 424 308]{Dehnen.fig1.ps}}
        \centerline{\fig{fuvobs}}
% FIGURE 5
	\centerline{\epsfxsize=75mm\epsfbox[7 6 169 354]{Dehnen.fig2.ps}}
        \centerline{\fig{simul}}

\fi	% !preprint
\end{document}